\documentstyle[aps,eqsecnum]{revtex}
\begin{document}
\draft
\twocolumn[\hsize\textwidth\columnwidth\hsize\csname 
           @twocolumnfalse\endcsname
\title{Radiative falloff in Schwarzschild-de Sitter spacetime}
\author{Patrick R.~Brady}
\address{Theoretical Astrophysics 130-33, California Institute of 
         Technology, Pasadena, California 91125\\
        Institute for Theoretical Physics,  University of California, Santa 
        Barbara CA 93106}
\author{Chris M.~Chambers$^*$}
\address{Department of Physics, Montana State University, 
         Bozeman, Montana 59717}
\author{William G.~Laarakkers and Eric Poisson}
\address{Department of Physics, University of Guelph, Guelph,
         Ontario, Canada N1G 2W1}
\date{Draft, January 29, 1999}
\maketitle
\begin{abstract}
We consider the evolution of a scalar field propagating in
Schwarzschild-de Sitter spacetime.  The field is non-minimally coupled
to curvature through a coupling constant $\xi$.  The spacetime has two
distinct time scales, $t_e = r_e/c$ and $t_c = r_c/c$, where $r_e$ is
the radius of the black-hole horizon, $r_c$ the radius of the
cosmological horizon, and $c$ the speed of light.  When $r_c \gg r_e$,
the field's time evolution can be separated into three epochs. At
times $t \ll t_c$, the field behaves as if it were in pure
Schwarzschild spacetime; the structure of spacetime far from the black
hole has no influence on the evolution.  In this early epoch, the
field's initial outburst is followed by quasi-normal oscillations, and
then by an inverse power-law decay.  At times $t \lesssim t_c$, the
power-law behavior gives way to a faster, exponential decay.  In this
intermediate epoch, the conditions at radii $r \gtrsim r_e$ and $r
\lesssim r_c$ both play an important role. Finally, at times $t \gg
t_c$, the field behaves as if it were in pure de Sitter spacetime; the
structure of spacetime near the black hole no longer influences the
evolution in a significant way.  In this late epoch, the field's
behavior depends on the value of the curvature-coupling constant
$\xi$. If $\xi$ is less than a critical value $\xi_c = 3/16$, the
field decays exponentially, with a decay constant that increases with
increasing $\xi$.  If $\xi > \xi_c$, the field oscillates with a
frequency that increases with increasing $\xi$; the amplitude of the
field still decays exponentially, but the decay constant is
independent of $\xi$. We establish these properties using a
combination of numerical and analytical methods.
\end{abstract}
\vskip 2pc]

\narrowtext

\section{Introduction and summary} 

The dynamics of radiative fields in black-hole spacetimes has been
investigated since the early 1970's in an effort to understand how
asymmetric gravitational collapse proceeds, in the absence of rotation, to
form a perfectly spherical black hole.  The first work of this type was
carried out by de la Cruz, Chase, and Israel \cite{1}, who focused on
electromagnetic and linear gravitational perturbations of a spherical
collapse. They showed numerically that after an outburst of radiation at
the onset of collapse, the radiation decays to zero in a time of the order
of the light-crossing time of the resulting event horizon.  Subsequently,
Price \cite{2} gave an analytical description of the radiative decay, and
showed that scalar, electromagnetic, and gravitational radiation all go to
zero as an inverse power of time, the power index depending on the
multipole order of the perturbation. This late-time decay of the radiation
is often referred to as its ``power-law tail''.

The early investigations were extended and improved by other
authors. Bi\v{c}\'ak \cite{3} generalized Price's analysis and calculated
the radiative decay of an electrically neutral scalar field interacting
with a charged black hole.  He also found an inverse power-law decay of the
radiation.  Leaver \cite{4} provided a detailed analytical picture of the
radiative evolution and introduced much of the mathematical framework
adopted in subsequent work.  Gundlach, Price, and Pullin~\cite{5} provided
the most complete picture to date; in addition to describing how the
radiation behaves at a fixed distance from the black hole, they also
described the behavior at the event horizon and at future null infinity.
Furthermore, they numerically simulated the collapse of a self-gravitating
scalar field, and showed that the power-law tails are a generic feature of
radiative decay, whether a black hole forms or not.  These numerical results
were reproduced by Burko and Ori~\cite{6}.  Additional analytical
insights can be found in Refs. \cite{7,8,9}, and various
generalizations of the basic scenario have been considered in \cite{10,11}.    

The works reviewed above were entirely concerned with gravitational
collapse to a non-rotating black hole. It is only recently that the
analysis was extended to the case of rotating black holes. The first
authors to consider radiative decay in Kerr spacetime were Krivan, Laguna,
and Papadopoulos \cite{12} (see also \cite{13}), who showed numerically
that the radiation falls off according to an inverse power law.  This
inverse power-law behavior was confirmed analytically by 
Hod \cite{14}, as well as Barak and Ori \cite{14.2} 

At this point, we have a clear physical picture of how a radiative field
behaves during gravitational collapse to a black hole; the picture is
essentially the same for scalar, electromagnetic, and gravitational
radiation, and it is valid for rotating and non-rotating black holes.
There are three stages to the field's dynamical evolution. At the onset of
collapse ($t=0$), an outburst of radiation is emitted; most of this
radiation propagates (with distortions) directly to infinity.  After this
initial outburst, the field oscillates with frequencies and damping times
characteristic of the central black hole.  This behavior is associated with
the field's quasi-normal modes of oscillation, which decay over a time
comparable to $100\, r_e/c$, where $r_e$ is the event-horizon radius and
$c$ the speed of light.  Finally, these oscillations give way to the
inverse power-law behavior described above. If the radiative field $\Phi$
is observed at a fixed radius $r$ and the field is static prior to
collapse, then $\Phi \sim t^{-(2l + 2)}$ as $t\rightarrow \infty$,
where $l$ is the multipole order.

Analytical~\cite{4,7,8,9} and numerical~\cite{5} studies of radiative
dynamics in black-hole spacetimes have revealed that the inverse power-law
behavior is not sensitive to the presence of an event horizon.
In fact, power-law tails are a weak-curvature
phenomenon, and it is the asymptotic structure of the spacetime at radii $r
\gg r_e$ which dictates how the field behaves at times $t \gg r_e/c$. This
observation begs the question: how is the field's evolution
affected if the conditions at infinity are altered?  For example, 
what happens when the black hole is immersed in an 
expanding universe \cite{15}?

This question was first addressed by Brady,
Chambers, Krivan, and Laguna (hereafter BCKL) in Ref.~\cite{16}. They
considered the dynamical evolution (both linear and nonlinear) of a
scalar field in Reissner-Norsdtr\"om-de Sitter spacetime, which
represents a spherical, charged black hole immersed in de Sitter 
space---an exponentially expanding universe with a cosmological
horizon at a fixed radius $r_c$. BCKL found that the scalar field
decays not as an inverse power of time, but exponentially at times $t
\gg r_c/c$. Their numerical results were compatible with the formula 
$\Phi \sim e^{-l\kappa_c t}$, where the decay constant $\kappa_c
\simeq c/r_c$ is the
surface gravity of the cosmological horizon. (The surface gravity is
precisely defined in Sec.~II.) For $l=0$, BCKL found
that the field does not decay, but that it settles down to a constant,
nonzero value. Therefore, the usual power-law scenario does not
necessarily survive when the conditions at infinity are altered.

Our purpose in this paper is to extend the initial work
of BCKL; we provide additional details, generalize their
discussion, and offer new analytical insights.  

We study the dynamics of a scalar field $\Phi$ in Schwarzschild-de
Sitter (SdS) spacetime, which describes an electrically neutral black
hole (of event-horizon radius $r_e$) immersed in an exponentially
expanding universe (with cosmological-horizon radius $r_c$). The
scalar field satisfies the wave equation
\begin{equation}
(\Box - \xi R) \Phi = 0,
\label{1.1}
\end{equation}
where $\Box$ is the curved spacetime d'Alembertian operator, $R$ the
spacetime's Ricci scalar, and $\xi$ a tunable, nonnegative coupling
constant. This equation is more general than the one considered by
BCKL, who limited themselves to minimal coupling, i.e., $\xi = 0$. We shall
see that adding this dimension to the parameter space greatly enhances
the number of possible late-time behaviors for the scalar
field. Additional details regarding the SdS spacetime and the wave
equation (\ref{1.1}) are presented in Sec.~II, which also describes
the numerical methods employed to integrate Eq.~(\ref{1.1}).

The analysis presented in this paper establishes the following physical
picture. The SdS spacetime comes with two distinct time scales, $r_e/c$ and
$r_c/c$; for the purpose of this discussion we assume that $r_e \ll r_c$ so
that the time scales are cleanly separated.  The picture elaborated
previously for black holes in asymptotically-flat spacetimes---an initial
outburst followed by quasi-normal oscillations followed by power-law
tails---continues to hold at times $t \ll r_c/c$. In fact, over this
time scale, the field propagates as if it were in asymptotically-flat
Schwarzschild spacetime.  Conditions at radii $r \simeq r_c$ have no
influence on the field's evolution. At later times, however, when $t$
becomes comparable to $r_c/c$, the asymptotic de Sitter structure starts
to play a role, and the inverse power-law behavior gives way to a faster,
exponential decay.  At times $t \gg r_c/c$ the conditions at radii $r
\gtrsim r_e$ no longer have any influence, and the scalar field goes
as $\Phi \sim e^{-p \kappa_c t}$, where
\begin{equation}
p = l + \frac{3}{2} - \frac{1}{2}\sqrt{9-48\xi} + 
O\biggl( \frac{r_e}{r_c} \biggr).
\label{1.2}
\end{equation}
Equation (\ref{1.2}) reduces to $p=l$ when $\xi = 0$, in agreement with the
BCKL results. In this expression,  notice that the square root
becomes imaginary when $\xi > \xi_c = \frac{3}{16}$. At $\xi = \xi_c$,
therefore, the scalar field undergoes a transition at which the pure
exponential decay observed for $\xi < \xi_c$ becomes oscillatory. This
surprising property of wave propagation in SdS spacetime was first
revealed in our numerical simulations (presented in Sec.~III), and
then established analytically (as presented in Sec.~IV). 

Our final section (Sec.~V) is devoted to a discussion of the late-time
decay of electromagnetic and gravitational radiation in SdS
spacetime. If $\Phi$ designates either one of these
radiative fields, we argue that $\Phi \sim e^{-p \kappa_c t}$ at late times,
where $p$ is now given by
\begin{equation}
p = l + 1 + O\biggl( \frac{r_e}{r_c} \biggr).
\label{1.3}
\end{equation}
Thus, both electromagnetic and gravitational radiation behave, at late
times, as a conformally invariant ($\xi = \frac{1}{6}$) scalar field. 

\section{Equations and numerical methods}

The metric of the Schwarzschild-de Sitter (SdS) spacetime is given by 
\begin{equation}
ds^2 = -f\, dt^2 + f^{-1}\, dr^2 + r^2 (d\theta^2 
+ \sin^2\theta\, d\phi^2), 
\label{2.1}
\end{equation}
where 
\begin{equation}
f = 1 - \frac{2M}{r} - \frac{r^2}{a^2},
\label{2.2}
\end{equation}
with $M$ denoting the black-hole mass; $a^2$ is given in terms of the
cosmological constant $\Lambda$ by $a^2 = 3/\Lambda$. We use
geometrized units such that $G=c=1$. The spacetime possesses two
horizons: the black-hole horizon is at $r=r_e$ and the
cosmological horizon is at $r = r_c$, where $r_c > r_e$. The function
$f$ has zeroes at $r_e$,  $r_c$, and $r_0 = -(r_e + r_c)$. In terms of
these quantities, $f$ can be expressed as
\begin{equation}
f = \frac{1}{a^2 r}\, (r-r_e)(r_c-r)(r-r_0).
\label{2.3}
\end{equation}
It is useful to regard $r_e$ and $r_c$ as the two fundamental
parameters of the SdS spacetime, and to express $M$ and $a^2$ as
functions of these variables. The appropriate relations are
\begin{equation}
a^2 = {r_e}^2 + r_e r_c + {r_c}^2
\label{2.4}
\end{equation}
and
\begin{equation}
2M a^2 = r_e r_c (r_e + r_c).
\label{2.5}
\end{equation}
We also introduce the surface gravity $\kappa_i$ associated with the
horizon $r = r_i$, as defined by the relation $\kappa_i = \frac{1}{2} 
|df/dr|_{r=r_i}$. Explicitly,  we have
\begin{eqnarray}
\kappa_e &=& \frac{ (r_c-r_e)(r_e-r_0) }{ 2a^2 r_e },
\label{2.6} \\
\kappa_c &=& \frac{ (r_c-r_e)(r_c-r_0) }{ 2a^2 r_c },
\label{2.7} \\
\kappa_0 &=& \frac{ (r_e-r_0)(r_c-r_0) }{ 2a^2 (-r_0) }.
\label{2.8}
\end{eqnarray}
These quantities allow us to write
\begin{equation}
\frac{1}{f} = \frac{1}{2\kappa_e (r-r_e)} +
\frac{1}{2\kappa_c (r_c-r)} +
\frac{1}{2\kappa_0 (r-r_0)},
\label{2.9}
\end{equation}
and to express the transformation between $r$ and the ``tortoise
coordinate'' $r^* \equiv \int f^{-1}\, dr$ as
\begin{eqnarray}
r^* &=& \frac{1}{2\kappa_e}\, \ln\biggl(\frac{r}{r_e} - 1\biggr) 
- \frac{1}{2\kappa_c} \ln\biggl(1 - \frac{r}{r_c}\biggr) 
\nonumber \\ & & \mbox{}
+ \frac{1}{2\kappa_0} \ln\biggl(\frac{r}{r_0} - 1\biggr).
\label{2.10}
\end{eqnarray}
In terms of $t$ and $r^*$ we define the null coordinates $u=t-r^*$
(retarded time) and $v=t+r^*$ (advanced time) so that the (future) black-hole
horizon is located at $u = \infty$,  and the (future) cosmological
horizon is at $v = \infty$. 

We consider a massless scalar field $\Phi$ in the SdS spacetime,
obeying the wave equation
\begin{equation}
( \Box - \xi R ) \Phi = 0,
\label{2.11}
\end{equation}
where $\Box = g^{\alpha\beta} \nabla_{\!\alpha} \nabla_{\!\beta}$ is
the d'Alembertian operator, $R = 12/a^2$ the Ricci scalar, and $\xi$ a
curvature-coupling constant that we take to be nonnegative. If we
decompose the scalar field according to 
\begin{equation}
\Phi = \sum_{lm} \frac{1}{r}\, \psi_l(t,r)\, Y_{lm}(\theta,\phi),
\label{2.12}
\end{equation}
then each wave function $\psi_l$ satisfies the equation
\begin{equation}
\biggl( -\frac{\partial^2}{\partial t^2} + 
\frac{\partial^2}{\partial r^{*2}} \biggr) \psi_l(t,r) 
= V_l(r) \psi_l(t,r),
\label{2.13}
\end{equation}
where the potential function is given by
\begin{equation}
V_l(r) = f \biggl[ \frac{l(l+1)}{r^2} + \frac{2M}{r^3} 
+ \frac{2(6\xi-1)}{a^2} \biggr].
\label{2.14}
\end{equation}
Using the null coordinates $u$ and $v$, Eq.~(\ref{2.13}) can be
recast as
\begin{equation}
-4 \frac{\partial^2}{\partial u \partial v}\, \psi_l(u,v)
= V_l(r) \psi_l(u,v),
\label{2.15}
\end{equation}
in which $r$ is determined by inverting the relation $r^*(r) =
\frac{1}{2}(v-u)$. Our numerical methods are based on this form for
the reduced wave equation.

In order to find a unique solution to Eq.~(\ref{2.15}),
initial data must be specified on the two null surfaces $u=0$ (say) and
$v=0$ (say). Because the late-time behavior of the wave function is
largely independent of the choice of initial data, we set $\psi_l(u,v=0) =
0$ and use a Gaussian profile,  centered on $v_c$ and having width
$\sigma$,  on $u=0$: 
\begin{equation}
\psi_l(u=0,v) = \exp\biggl[-\frac{(v-v_c)^2}{2\sigma^2} \biggr].
\label{2.16}
\end{equation}

We numerically integrate Eq.~(\ref{2.15}) using the finite-differencing
scheme suggested by Gundlach, Price, and Pullin \cite{5}, in which the 
coordinates $u$ and $v$ increase by discrete units $\Delta$. In the
discrete space, the differential equation becomes
\begin{eqnarray}
\psi_l(N) &=& \psi_l(W) + \psi_l(E) - \psi(S) 
\nonumber \\ & & \mbox{}
- \frac{\Delta^2}{8}\, V_l(R_c) \Bigl[ \psi_l(W) + \psi_l(E) \Bigr] 
+ O(\Delta^4),
\label{2.17}
\end{eqnarray}
where we have defined the points $N: (u+\Delta,v+\Delta)$, 
$W: (u+\Delta,v)$, $E: (u, v+\Delta)$, and $S: (u,v)$.
The potential is evaluated at the central radius $R_c$
corresponding to the off-grid point $(u+\frac{1}{2}\Delta,
v+\frac{1}{2}\Delta)$; thus, $r^*(R_c) = \frac{1}{2}(v-u)$. 

The computationally expensive part of the calculation resides in the
inversion of the relation $r^*(r)$, which is required to evaluate the
potential function $V_l(r)$. For values of $r^*$ such that $2 \kappa_e r^*
< -1$, we numerically solve the equation
\begin{equation}
\frac{r}{r_e} = 1 + e^{2\kappa_e r^*} \exp\Biggl[ 
\frac{\kappa_e}{\kappa_c} \ln\biggl(1-\frac{r}{r_c}\biggr) - 
\frac{\kappa_e}{\kappa_0} \ln\biggl(1-\frac{r}{r_0}\biggr) \Biggr]
\label{2.18}
\end{equation}
by iterations. When $2\kappa_c r^* > 1$, we use
\begin{equation}
\frac{r}{r_c} = 1 - e^{-2\kappa_c r^*} \exp\Biggl[ 
\frac{\kappa_c}{\kappa_e} \ln\biggl(\frac{r}{r_e}-1\biggr) + 
\frac{\kappa_c}{\kappa_0} \ln\biggl(1-\frac{r}{r_0}\biggr) \Biggr]
\label{2.19}
\end{equation}
instead. For all other values of $r^*$, we invert the
relation $r^*(r)$ using the Newton-Raphson method, as implemented in
the {\it Numerical Recipes} routine {\tt rtsafe} \cite{17}. 

Once the integration is completed, the values $\psi_l(u_{\rm max},v)$
are extracted, where $u_{\rm max}$ is the maximum value of $u$ on the
numerical grid. If $u_{\rm max}$ is sufficiently large, then these
values give a good approximation for the wave function at the event
horizon. Similarly, for $v_{\rm max}$ sufficiently large, the values
$\psi_l(u,v_{\rm max})$ give a good approximation for the wave
function on the cosmological horizon. Finally, the values of
$\psi_l$ on the line $u = v - 2k$ are extracted, and expressed as a
function of $t = \frac{1}{2}(u+v) = v-k$; here, $k$ is a constant
representing the value of $r^*$ at which the field is evaluated.  

\section{Numerical results}

\subsection{Behavior at early and intermediate times}

Our first goal in this section is to establish that at times $t
\ll r_c$, the scalar field behaves as if it were in Schwarzschild
spacetime, and that at times $t \sim r_c$, a transition
occurs in which the inverse power-law decay gives way to an
exponential decay. The late times $t \gg r_c$ are considered in
the next subsection.  

Figure 1 displays the behavior of the wave function $\psi_l(t,r)$ at a
fixed radius $r$, for early and intermediate times. The plots show how the
pure Schwarzschild behavior (initial outburst, quasi-normal oscillations,
and power-law decay) is distorted by the de Sitter structure of spacetime
at large radii.  We consider a SdS spacetime
with parameters $r_e = 1$ and $r_c = 2000$, so that the time scales
are well separated. We integrate Eq.~(\ref{2.15}) for $l=0$ and $l=1$,
setting $\xi = 0$. We compare the field's evolution in SdS spacetime
to what it would be in pure Schwarzschild spacetime (also with
$r_e = 1$). In both cases the initial data is given by
Eq.~(\ref{2.16}), with $v_c = 10$, $\sigma = 3$, and the wave function
is evaluated at $r^* = 10$. 

\begin{figure}
\special{hscale=35 vscale=35 hoffset=-25.0 voffset=10.0
         angle=-90.0 psfile=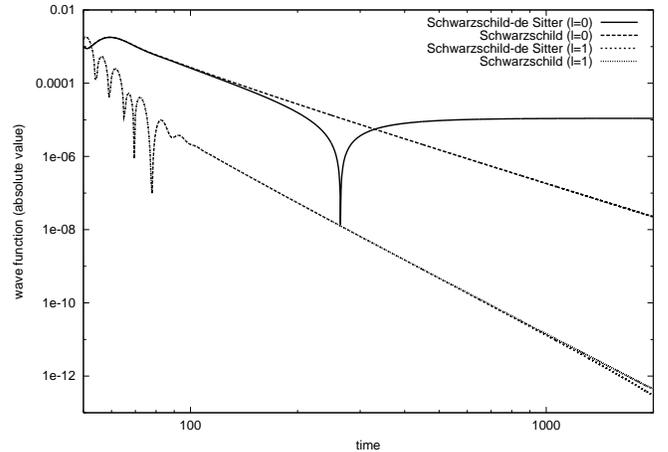}
\vspace*{2.6in}
\caption{Absolute value of the wave function $\psi_l(t,r)$ as a
function of time $t$, evaluated at $r^* = 10$ in Schwarzschild
spacetime ($r_e = 1$) and SdS spacetime ($r_e = 1$ and $r_c =
2000$). The cases $l=0$ and $l=1$ are considered, and the wave
functions are plotted on a log-log scale. In such a plot, a straight
line indicates power-law behavior, and a change of sign in the wave
function is represented by a deep trough. We see that the early
portion of $\psi_1$ is oscillatory, and that for SdS spacetime,
$\psi_0$ changes sign at $t \sim 260$.} 
\end{figure}

The plots show that the physical picture given in Sec.~I is
accurate. At early times, the Schwarzschild and SdS wave functions
display identical behavior, and deviations become apparent only when
$t$ becomes comparable to $r_c$. For $l=0$, the
Schwarzschild behavior $\psi_0 \sim t^{-3}$ gives way to the wave
function changing sign at $t \sim 260$, and settling down to a
constant value (equal to $-1.1 \times 10^{-5}$) at late times. For
$l=1$, the Schwarzschild behavior $\psi_1 \sim t^{-5}$ is replaced by
a faster decay, which eventually becomes exponential. Examination of
the field's behavior at the black-hole and
cosmological horizons, where $\psi_l$ is a function of $u$ and $v$,
respectively, reveals precisely the same physical picture. 

\subsection{Behavior at late times}

We now consider the behavior of the scalar field at times $t \gg
r_c$. To ease access to these late times, we consider a SdS spacetime
with parameters $r_e = 1$ and $r_c = 100$, although this no longer
offers a clean separation of the time scales. We begin by integrating
Eq.~(\ref{2.15}) with $\xi = 0$, for the cases $l=0,1,2$. Once again
we set $v_c = 10$, $\sigma = 3$, and we evaluate the wave function at
$r^* = 10$. 

Figure 2 displays the behavior of the wave function $\psi_l(t,r)$ at
late times. The plots reveal that for $l=0$, the wave function
eventually settles down to the constant value $-4.4 \times 10^{-3}$.
For $l \neq 0$, the field decays to zero, and our numerical results
are compatible with the formula
\begin{equation}
\psi_l(t) \sim  e^{-l\kappa_c t}
\label{3.1}
\end{equation}
first discovered by BCKL.  The numerically-determined values for the
decay constants agree with $l\kappa_c$ to within 1\%. Multipoles
corresponding to $l \geq 3$ could not be examined because the field 
decays too rapidly and the numerical integration
quickly becomes noisy. The late-time behavior of the wave function at
the black-hole and cosmological horizons is also described by
Eq.~(\ref{3.1}), provided that $t$ is replaced by $v$ on the event
horizon, and by $u$ on the cosmological horizon.

\begin{figure}
\special{hscale=35 vscale=35 hoffset=-25.0 voffset=10.0
         angle=-90.0 psfile=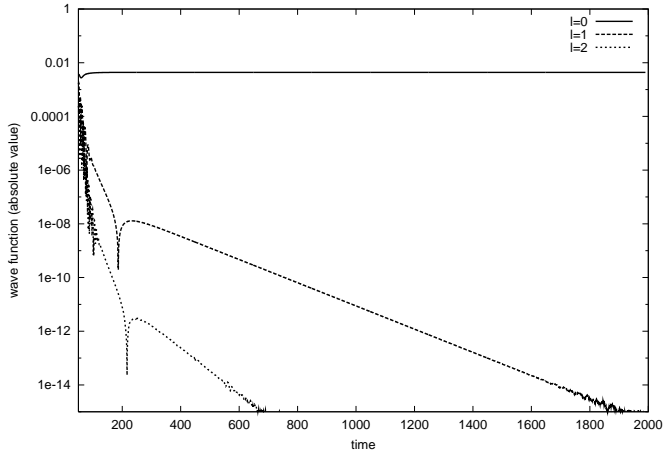}
\vspace*{2.6in}
\caption{Absolute value of the wave function $\psi_l(t,r)$ as a
function of time $t$, evaluated at $r^* = 10$ in SdS spacetime 
($r_e = 1$ and $r_c = 100$). The cases $l=0, 1, 2$ are considered, and
the wave functions are plotted on a semi-log scale. In such a plot, a
straight line indicates exponential behavior. Notice that the final
change of sign of the wave function occurs at $t \sim 50$ for $l=0$, 
$t \sim 190$ for $l=1$, and $t \sim 220$ for $l=2$. Notice also that
the numerical integration becomes noisy when $|\psi_l|$ drops below 
$10^{-14}$.} 
\end{figure}

A rich spectrum of late-time behaviors is revealed when $\xi$, the
curvature-coupling constant, is allowed to be nonzero. This dimension
of the parameter space was not explored by BCKL. Figure 3 displays
the late-time behavior of $\psi_0$ for several values of $\xi$, using
the same parameters as before. For $\xi$ smaller than a critical value
$\xi_c$, the field decays
monotonically with a decay constant that increases with increasing
$\xi$. When $\xi > \xi_c$, however, the wave function oscillates
with a decaying amplitude. As $\xi$ is increased away
from the critical value $\xi_c$, the frequency of the oscillations
increases, but the decay constant stays the same.  Similar behavior
was observed for $l=1$ and $l=2$, and our numerical results
are compatible with the formula
\begin{equation}
\psi \sim e^{-[l + g(\xi)] \kappa_c t},
\label{3.2}
\end{equation}
where the function $g(\xi)$ increases with $\xi$ for $\xi < \xi_c$, and is
complex for $\xi > \xi_c$. From our numerical results, we are able to
determine that $0.18 \lesssim \xi_c \lesssim 0.19$. The analytical
calculations presented in Sec. IV show that the function $g(\xi)$ is in
fact given by
\begin{equation}
g(\xi) = \frac{3}{2} - \frac{1}{2} \sqrt{9-48\xi} + 
O\biggl( \frac{r_e}{r_c} \biggr),
\label{3.3}
\end{equation}
so that $\xi_c = \frac{3}{16} = 0.1875$. This relation explains the
observed features, and our numerical results agree with
Eqs.~(\ref{3.2}) and (\ref{3.3}) to within 1\%. Equation (\ref{3.2}),
with $t$ replaced by either $v$ or $u$, describes also the behavior of
the wave function on the black-hole or cosmological horizons,
respectively.

\begin{figure}
\special{hscale=35 vscale=35 hoffset=-25.0 voffset=10.0
         angle=-90.0 psfile=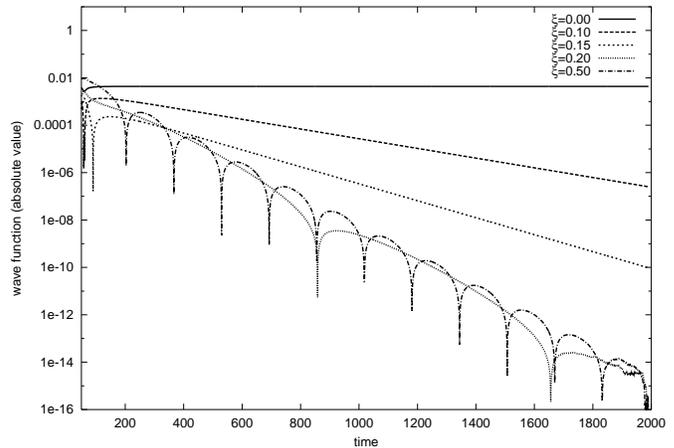}
\vspace*{2.6in}
\caption{Absolute value of the wave function $\psi_0(t,r)$ as a
function of time $t$, evaluated at $r^* = 10$ in SdS spacetime 
($r_e = 1$ and $r_c = 100$). Several values of $\xi$ are considered,
in the interval between $\xi = 0$ and $\xi = \frac{1}{2}$. The wave
functions are plotted on a semi-log scale. The noteworthy features are
these: (i) For $\xi < \xi_c$, the wave function decays exponentially,
with a decay constant that increases with increasing $\xi$; (ii) for
$\xi > \xi_c$, the wave function still decays exponentially, but with
a decay constant that no longer varies with $\xi$; (iii) for $\xi >
\xi_c$, the wave function oscillates, with a frequency that increases
with $\xi$.} 
\end{figure}

According to the physical picture presented in Sec.~I, the field's
evolution should proceed as if the field were in pure de Sitter
spacetime when $t \gg r_c$.  In particular, Eqs.~(\ref{3.2}) and
(\ref{3.3}) should describe the late-time behavior of
the wave function in de Sitter spacetime. To test this hypothesis, we 
performed a numerical integration of Eq.~(\ref{2.15}) in pure de
Sitter spacetime \cite{XX} with a cosmological horizon at $r_c = 1$, and 
initial data of the form (\ref{2.16}) with $v_c = 1$ and $\sigma
= 0.1$. The results for $l=0$
are displayed in Figure 4.  All the late-time features
seen in SdS spacetime are reproduced, and we have verified that the
field's evolution is well described by
Eqs.~(\ref{3.2}) and (\ref{3.3}), with $r_e = 0$ and 
$\kappa_c = 1/r_c = 1$, to well within 1\%.

\begin{figure}
\special{hscale=35 vscale=35 hoffset=-25.0 voffset=10.0
         angle=-90.0 psfile=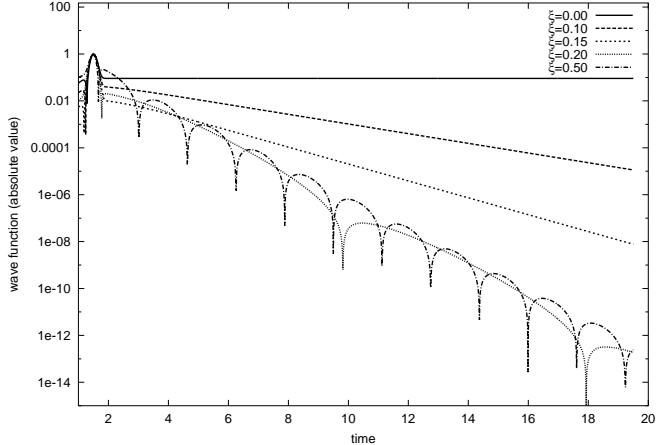}
\vspace*{2.6in}
\caption{Absolute value of the wave function $\psi_0(t,r)$ as a
function of time $t$, evaluated at $r^* = 0.5$ in pure de Sitter
spacetime (with parameter $r_c = 1$). Several values of $\xi$ are
considered, in the interval between $\xi = 0$ and $\xi =
\frac{1}{2}$. The wave functions are plotted on a semi-log scale. 
These plots show the same features as in the preceding figure.} 
\end{figure}

\section{Analytical results}

The observation that the late-time behavior of the scalar
field is dictated by the de Sitter structure of spacetime at large
radii is the key to the derivation of Eqs.~(\ref{3.2}) and
(\ref{3.3}). Motivated by this observation, we integrate the wave
equation (\ref{2.13}) in pure de Sitter spacetime, and show that the
late-time behavior of the wave function is indeed given by
Eq.~(\ref{3.2}), with $g(\xi) = \frac{3}{2} - \frac{1}{2}
\sqrt{9-48\xi}$. Apart from an approximation of late times ($t \gg
r_c$), our calculations are {\it exact}.  That such a
treatment is afforded relies on removing the black
hole from the spacetime. While this removal produces a large effect at
early times, the preceding results indicate that the influence of the
central mass is negligible at late times. (We will have to qualify
this statement below.) The considerably more difficult job of
analytically integrating Eq.~(\ref{3.2}) in full SdS spacetime will be
undertaken elsewhere \cite{18}.

\subsection{Equations}

We consider a scalar field $\Phi$ in de Sitter spacetime, whose 
metric is of the form (\ref{2.1}) with
\begin{equation}
f = 1 - r^2,
\label{4.1}
\end{equation}
where we have set $r_c = a = 1$ without loss of generality; thus, the
surface gravity of the cosmological horizon is
$\kappa_c = 1$. The scalar field obeys the wave equation (\ref{2.11}),
and the reduced wave function $\psi(t,r)$ defined by Eq.~(\ref{2.12})
satisfies Eq.~(\ref{2.13}) with the potential
\begin{equation}
V(r) = f \biggl[ \frac{l(l+1)}{r^2} + 2(6\xi - 1) \biggr].
\label{4.2}
\end{equation}
We suppress the $l$-label to simplify the notation. In de
Sitter spacetime, the relation between $r$ and $r^* \equiv \int
f^{-1}\, dr$ is given by
\begin{equation}
e^{2r^*} = \frac{1+r}{1-r}.
\label{4.3}
\end{equation}
We wish to integrate Eq.~(\ref{2.13}) starting with a suitable set of
initial conditions at $t=0$. It follows from a simple application of
Green's identity that at times $t > 0$, the wave function can be
expressed as
\begin{equation}
\psi(t,r) = - \int_0^1 \Bigl[ \psi(0,r') \dot{g}(t;r,r') 
+ \dot{\psi}(0,r') g(t;r,r') \Bigr]\, \frac{dr'}{f'}.
\label{4.4}
\end{equation}
Here, $\psi(0,r')$ and $\dot{\psi}(0,r')$ are the initial data (an
overdot indicates time differentiation \cite{foot1}), and $g(t;r,r')$
is the retarded Green's function for the reduced wave equation
(\ref{2.13}); we also use the notation $f' \equiv 1 - r'^2$. The
Green's function is zero for $t < 0$, and it satisfies
Eq.~(\ref{2.13}) with a term $f \delta(t) \delta(r-r')$ added to the
right-hand side.

\subsection{Fourier transform of the Green's function}

Since the late-time behavior of $\psi(t,r)$ is entirely determined
by the late-time behavior of $g(t;r,r')$, the derivation of
Eqs.~(\ref{3.2}) and (\ref{3.3}) begins with a calculation of
the Green's function. We first consider its Fourier transform 
$\tilde{g}(\omega;r,r')$, defined by 
\begin{equation}
g(t;r,r') = \frac{1}{2\pi}\, \int \tilde{g}(\omega;r,r') 
e^{-i\omega t}\, d\omega.
\label{4.5}
\end{equation}
For each $\omega$, this function of $r$ and $r'$ satisfies the
differential equation
\begin{equation}
\biggl[ \frac{d^2}{dr^{*2}} + \omega^2 - V(r) \biggr] 
\tilde{g}(\omega;r,r') = f \delta(r-r'),
\label{4.6}
\end{equation}
where $V(r)$ is given by Eq.~(\ref{4.2}). It is clear that the
solution can be expressed in the factorized form
\begin{equation}
\tilde{g}(\omega;r,r') = \frac{1}{W}\, 
\psi^{\rm reg}(r_<) \psi^{\rm up}(r_>),
\label{4.7}
\end{equation}
where $\psi^{\rm reg}(r)$ and $\psi^{\rm up}(r)$ are two linearly
independent solutions to the homogeneous form of Eq.~(\ref{4.6}),
with $W$ denoting their conserved Wronskian:
\begin{equation}
W = \psi^{\rm reg} \frac{d}{dr^*} \psi^{\rm up} - 
\psi^{\rm up} \frac{d}{dr^*} \psi^{\rm reg}.
\label{4.8}
\end{equation}
We use the notation $r_< = \mbox{min}(r,r')$, $r_> =
\mbox{max}(r,r')$, and for simplicity, we do not explicitly indicate
the dependence of $\psi^{\rm reg}$, $\psi^{\rm up}$, and $W$ on
$\omega$. 

Equation (\ref{4.7}) correctly represents the retarded Green's
function if $\psi^{\rm up}(r)$ describes waves which are purely
outgoing at the cosmological horizon; we choose the normalization 
\begin{equation}
\psi^{\rm up} \sim e^{i\omega r^*} \qquad r \to 1.
\label{4.9}
\end{equation}
Also, we must choose for $\psi^{\rm reg}(r)$ a solution which enforces 
the boundary condition $\psi^{\rm reg}(0) = 0$;
this ensures that the scalar field $\Phi$ is everywhere
nonsingular. The calculation of $\tilde{g}(\omega;r,r')$ therefore
reduces to solving the homogeneous form of Eq.~(\ref{4.6}) for the two
functions $\psi^{\rm reg}(r)$ and $\psi^{\rm up}(r)$.

\subsection{Solutions to the homogeneous equation}

The mathematical problem of finding the solutions $\psi^{\rm reg}(r)$
and $\psi^{\rm up}(r)$ has been addressed in the literature
\cite{19,20}. We shall briefly sketch the method of solution. 

The change of variable $z=r^2$ and the factorization 
$\psi = z^\alpha (1-z)^\beta X$, with $\alpha =
\frac{1}{2} (l+1)$ and $\beta = \pm \frac{i}{2} \omega$, transforms
the homogeneous form of Eq.~(\ref{4.6}) to a hypergeometric equation
for $X$. The fundamental solutions are $X =
F(a_\pm,b_\pm;l+\frac{3}{2};z)$, where $a_\pm = \frac{1}{2}(l +
g_+ \pm i\omega)$, $b_\pm = \frac{1}{2}(l + g_- \pm i\omega)$, and 
\begin{equation}
g_\pm(\xi) = \frac{3}{2} \pm \frac{1}{2}\sqrt{9 - 48\xi}.
\label{4.11}
\end{equation}
Both choices of sign in front of $i\omega$ produce a function
$\psi(r)$ that goes to zero (as $r^{l+1}$) when $r \to 0$. Choosing
the minus sign and the normalization arbitrarily, we set
\begin{equation}
\psi^{\rm reg}(r) = r^{l+1}(1-r^2)^{-i\omega/2} 
F(a_-,b_-;l+{\textstyle \frac{3}{2}};r^2).
\label{4.12}
\end{equation}
It is not difficult to show that
\begin{equation}
\psi^{\rm up}(r) = 2^{i\omega} r^{l+1}(1-r^2)^{-i\omega/2} 
F(a_-,b_-;1-i\omega;1-r^2)
\label{4.13}
\end{equation}
also satisfies the homogeneous form of Eq.~(\ref{4.6}), and has the
asymptotic behavior indicated in Eq.~(\ref{4.9}). On the other hand,
the solution
\begin{equation}
\psi^{\rm down}(r) = 2^{-i\omega} r^{l+1}(1-r^2)^{i\omega/2} 
F(a_+,b_+;1+i\omega;1-r^2)
\label{4.14}
\end{equation}
satisfies ingoing-wave boundary conditions at the cosmological
horizon: $\psi^{\rm down}(r) \sim e^{-i\omega r^*}$ when $r \to 1$. 

These three solutions are not linearly independent. Using Eq.~(15.3.6)
of Abramowitz and Stegun \cite{21}, it is easy to show that
\begin{eqnarray}
\psi^{\rm reg}(r) &=& \frac{2^{-i\omega} \Gamma(l+\frac{3}{2}) 
\Gamma(i\omega)}{\Gamma(a_+) \Gamma(b_+)}\, \psi^{\rm up}(r) 
\nonumber \\ & & \mbox{}
+ \frac{2^{i\omega} \Gamma(l+\frac{3}{2}) 
\Gamma(-i\omega)}{\Gamma(a_-) \Gamma(b_-)}\, \psi^{\rm down}(r).
\label{4.15}
\end{eqnarray}
This relation, together with the fact that the Wronskian between 
$\psi^{\rm down}(r)$ and $\psi^{\rm up}(r)$ is equal to $2i\omega$,
imply that  
\begin{equation}
\frac{1}{W} = - \frac{1}{2^{1+i\omega}}\,
\frac{\Gamma\bigl[\frac{1}{2}(l+g_+ - i\omega)\bigr] 
\Gamma\bigl[\frac{1}{2}(l+g_- - i\omega)\bigr]}
{\Gamma(l+\frac{3}{2}) \Gamma(1 -i\omega)}.
\label{4.16}
\end{equation}

\subsection{Pole structure}

If for the moment we ignore the factor $1/\Gamma(1-i\omega)$,
Eq.~(\ref{4.16}) shows that $1/W$ [or equivalently
$\tilde{g}(\omega;r,r')$] is analytic in the complex $\omega$
plane, except for isolated poles in the lower half plane, at 
\begin{equation}
i\omega = i\omega^\pm_n \equiv l + g_{\pm} + 2n,
\label{4.17}
\end{equation}
where $g_{\pm}(\xi)$ is defined in Eq.~(\ref{4.11}) and $n =
0,1,2,\cdots$. There are two poles for each value of the integer $n$. 
For $\xi < \xi_c \equiv \frac{3}{16}$, $g_{\pm}$ are real functions of
$\xi$, and the poles are located along the negative imaginary axis of
the complex $\omega$ plane. As $\xi \to \xi_c$, the single poles
belonging to the same $n$ merge, becoming a double pole when $\xi =
\xi_c$. For $\xi > \xi_c$, the functions $g_{\pm}$ become complex, and
the poles move away from the negative imaginary axis.

The pole nearest to the real axis gives the dominant contribution to
the Green's function at late times.  For $\xi < \xi_c$, this pole is
located at $i\omega = l + \frac{3}{2} - \frac{1}{2}\sqrt{9 -
48\xi}$. For $\xi = \xi_c$, the double pole is located at $i\omega = l
+ \frac{3}{2}$. For $\xi > \xi_c$, there are two nearest poles at 
$i\omega = l + \frac{3}{2} \pm \frac{i}{2}\sqrt{48\xi - 9}$. 

For values of $\xi$ such that $\sqrt{9-48\xi}$ is an odd integer, the
numbers $\omega_n^\pm$ are also integers. This occurs when $\xi = 0$,
for which the poles are at $i\omega = l + 2n$ and $i\omega = l + 3
+2n$. It also occurs when $\xi = \frac{1}{6}$ (the value of $\xi$
which makes the wave equation conformally invariant), for which the
poles are at $i\omega = l + 1 + 2n$ and $i\omega = l + 2 +
2n$.  These are exceptional cases, because most of these poles are
canceled out by the previously ignored factor
$1/\Gamma(1-i\omega)$. This $\Gamma$-function also has integer
poles, at $i\omega = 1 + n'$, where $n'=0,1,2,\cdots$.  It is easy to
check that the zeros of $1/\Gamma(1-i\omega)$ cancel out the
poles listed previously, so that $1/W$ [or equivalently, 
$\tilde{g}(\omega;r,r')$] is analytic at these
frequencies. The only exception occurs for $l=0$ when $\xi = 0$. In
this case, the single pole at $i\omega = 0$ survives.  This is why the 
scalar field settles down to a nonzero constant at late times. 

The pole cancellation which occurs at integer values of $i\omega$ has
interesting consequences. For example, consider the case $l=1$, and values
of $\xi$ which are small but nonzero. For these values, the pole nearest to
the real axis is at $i\omega = 1 + 4\xi + O(\xi^2)$, and this pole exists
as long as $\xi \neq 0$. In the $\xi \rightarrow 0$ limit, however, the
would-be pole at $i\omega = 1$ is canceled by the compensating zero
in $1/\Gamma(1-i\omega)$.  We should therefore expect the Green's 
function---and the scalar field---to display a {\it qualitative}
change of behavior at late times, as the limit $\xi = 0$ is taken. Figure 5
shows that such is indeed the case.  It is interesting to note that such
qualitative changes of behavior do {\it not} occur in full SdS
spacetime. We return to this point in Sec.~IV G.

\begin{figure}
\special{hscale=35 vscale=35 hoffset=-25.0 voffset=10.0
         angle=-90.0 psfile=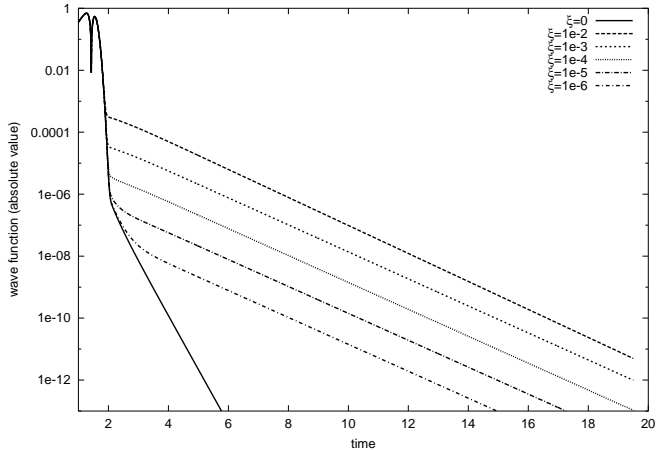}
\vspace*{2.6in}
\caption{Absolute value of the wave function $\psi_1(t,r)$ as a
function of time $t$, evaluated at $r^* = 0.5$ in pure de Sitter
spacetime (with parameter $r_c = 1$). Several small values of $\xi$ 
are considered, together with the special case $\xi = 0$. The plots
make it quite clear that the late-time behavior associated with 
$\xi \ll 1$ is qualitatively different from the behavior associated
with $\xi = 0$. As explained in the text, this qualitative change of
behavior is caused by pole cancellation in $1/W$.}
\end{figure}

The presence of poles in $\tilde{g}(\omega;r,r')$
makes it very easy to calculate the late-time behavior of the Green's
function. Such a calculation is much more difficult when there are no
poles, and therefore, this will not be attempted for the special cases 
$\xi = 0$ (except for $l=0$) and $\xi = \frac{1}{6}$. In the generic
situation (all other cases), the contribution to the Green's function
coming from the pole nearest to the real axis is given by
\begin{equation}
\tilde{g}(\omega;r,r') \sim 
\frac{\Gamma\bigl(\frac{1}{2}\sqrt{9-48\xi}\bigr)}
{\Gamma(l+\frac{3}{2}) \Gamma(1-l-g_-)}\, 
\frac{h(r) h(r')}{i(\omega - \omega_0)},
\label{4.18}
\end{equation}
where
\begin{equation}
h(r) = r^{l+1} (1-r^2)^{-(l+g_-)/2}
\label{4.18.5}
\end{equation}
and 
\begin{equation}
i\omega_0 = l + \frac{3}{2} - \frac{1}{2}\sqrt{9-48\xi}.
\label{4.19}
\end{equation}
This calculation involves substituting the asymptotic relation
$\Gamma(\epsilon) \sim 1/\epsilon$ in place of the relevant
$\Gamma$-function, and evaluating every other factor comprising 
$\tilde{g}(\omega;r,r')$ at the frequency $\omega_0$;
Eqs.~(\ref{4.7}), (\ref{4.12}), (\ref{4.13}), and (\ref{4.16}) are
used along the way \cite{foot2}. When $\xi = l = 0$, 
Eqs.~(\ref{4.18})--(\ref{4.19}) reduce to
\begin{equation}
\tilde{g}(\omega;r,r') \sim \frac{r r'}{i\omega}.
\label{4.20}
\end{equation}

\subsection{Green's function at late times}

The Green's function $g(t;r,r')$ can be obtained from
$\tilde{g}(\omega;r,r')$ by inverting the Fourier transform; this
gives rise to Eq.~(\ref{4.5}). The contour of integration is 
along the real axis of the complex $\omega$ plane.  However, because we
are interested in the behavior of the Green's function at late times, we
can safely close the contour in the lower half plane without changing
the result of the integral. The residue theorem then guarantees that
$g(t,r,r')$ can be expressed as a sum over residues, each pole
contributing a term proportional to $\exp(-i\omega^\pm_{n} t)$ to the
Green's function.  The dominant contribution comes from the
pole nearest to the real axis, and we obtain
\begin{equation}
g(t;r,r') \sim - \frac{\Gamma\bigl(\frac{1}{2}\sqrt{9-48\xi}\bigr)}
{\Gamma(l+\frac{3}{2}) \Gamma(1-l-g_-)}\, h(r) h(r')\, e^{-(l + g_-)t},
\label{4.21}
\end{equation}
where $g_-(\xi) = \frac{3}{2} - \frac{1}{2}\sqrt{9-48\xi}$. Near the
cosmological horizon, $1-r^2 \simeq 4 e^{-2 r^*}$, 
and Eq.~(\ref{4.21}) becomes
\begin{eqnarray}
g(t;r\to 1,r') &\sim& - 2^{-(l+g_-)}\, 
\frac{\Gamma\bigl(\frac{1}{2}\sqrt{9-48\xi}\bigr)} 
{\Gamma(l+\frac{3}{2}) \Gamma(1-l-g_-)}
\nonumber \\ & & \mbox{} \times
h(r')\, e^{-(l + g_-)u},
\label{4.22}
\end{eqnarray}
where $u = t - r^*$. For $\xi = l = 0$, both expressions reduce to
\begin{equation}
g(t;r,r') \sim - r r'.
\label{4.23}
\end{equation}
Equations (\ref{4.21})--(\ref{4.23}) are valid at
times $t \gg 1$, or $u \gg 1$. 

We have found that in the generic situation, the late-time behavior of
the Green's function is given by $e^{-p\kappa_c t}$ or 
$e^{-p\kappa_c u}$, where  
\begin{equation}
p = l + \frac{3}{2} - \frac{1}{2}\sqrt{9-16\xi}.
\label{4.24}
\end{equation}
Notice that we have re-introduced the scale parameter $\kappa_c =
1/r_c$. For the special case $\xi = l = 0$, the decay constant
$p$ is zero, and the Green's function settles down to the 
time-independent expression given by Eq.~(\ref{4.23}).

Equation (\ref{4.24}) gives the correct late-time behavior of the
Green's function for $\xi < \xi_c$ and $\xi > \xi_c$. For $\xi =
\xi_c$, the Green's function is obtained by integrating around a
double pole at $i\omega = l + \frac{3}{2}$. This requires integration
by parts, and the end result is that for fixed $r < r_c$, the Green's
function behaves as $(\kappa_c t) \exp[-(l+\frac{3}{2})\kappa_c t]$;
with $u$ replacing $t$, this is also the correct behavior on the
cosmological horizon. 

By virtue of Eq.~(\ref{4.4}), we conclude that in pure de Sitter
spacetime, the late-time behavior of the wave function $\psi(t,r)$
is also $e^{-p\kappa_c t}$ or $e^{-p\kappa_c u}$, except for the
exceptional cases---$\xi = 0$ and $\xi = \frac{1}{6}$---discussed
previously. For the special case $\xi=l=0$, the wave function settles
down to a constant value. This conclusion
is in full accord with the numerical results presented in Figs.~4 and
5.  

\subsection{The special case $\xi = l = 0$}

When $\xi = l = 0$, the wave function settles down to a
final value $\psi(\infty,r)$ which can readily be calculated using
Eqs.~(\ref{4.4}) and (\ref{4.23}). Re-introducing the scale factor
$r_c$, we obtain
\begin{equation}
\psi(\infty,r) = \frac{r}{{r_c}^2}\, \int_0^{r_c} \dot{\psi}(0,r')\,
r'\, \frac{dr'}{f'}.
\label{4.25}
\end{equation}
This equation implies that the value of the scalar field
$\Phi$ at late times does not vary with $r$, a necessary consequence
of the cosmological no-hair theorems \cite{22,23,24}. It also implies
that $\psi(\infty,r)$ scales as $1/{r_c}^2 = \Lambda/3$, a property
that was discussed in BCKL.

Equation (\ref{4.25}) relates $\psi(\infty,r)$ to the initial data
specified on the spacelike surface $t=0$. We can extract more
information from this equation, and also test it against our numerical
results, if we instead express $\psi(\infty,r)$ in terms of initial
data specified on the null surface $u=0$. (We again assume zero data
on $v=0$.) To effect this translation, we assume that the initial data
has support solely in the ``weak-field'' region of de Sitter
spacetime, in which $(r/r_c)^2 \ll 1$, so that $f \simeq 1$ and $r^*
\simeq r$. In this region, and in a neighborhood of the surface $t=0$,
we can approximate $\psi(t,r)$ by the flat-spacetime solution $\psi(t,r)
\simeq H(v)$, where $H$ is an arbitrary function of $v = t + r$. For
this solution, $\dot{\psi}(0,r) \simeq H'(r)$ (with a prime indicating 
differentiation with respect to the argument), and the integral in
Eq.~(\ref{4.25}) becomes $\int r dH$. After integration by parts,
assuming that $H(v)$ has compact support, we obtain $-\int H(r)\, dr$,
and 
\begin{equation}
\psi(\infty,r) \simeq - \frac{r}{{r_c}^2}\, \int_0^\infty H(v)\, dv.
\label{4.26}
\end{equation}
Recall that $H(v)$ is the initial wave profile on the surface
$u=0$. For the specific choice (\ref{2.16}), we obtain
\begin{equation}
\psi(\infty,r) \simeq -\frac{\sqrt{2\pi}\sigma}{{r_c}^2}\, r.
\label{4.27}
\end{equation}
We have tested this prediction against our numerical results. The
accuracy of Eq.~(\ref{4.27}) depends strongly on the value of
$v_c$, the central position of the Gaussian wave packet.  If $r_c =
1$, we find that for $v_c < 0.2$, our expression is
accurate to within 1\% over a wide selection of values for $r$ and
$\sigma$. For $v_c > 0.2$, the accuracy gets increasingly worse as
$v_c$ increases; for example, the error grows to 15\% at $v_c =
0.8$. In view of the fact that Eq.~(\ref{4.27}) was derived under the
assumption that the initial data has support only in the region 
$r \ll r_c$ of de Sitter spacetime, this is the expected result. 

The rate at which the field $\psi$ settles down to its final constant
value was also explored by BCKL. Their numerical results indicated
that $\psi - \psi(\infty,r) \sim e^{-q \kappa_c t}$ where $q \simeq
2$. Inspection of Eqs.~(\ref{4.11}) and (\ref{4.17}) reveals that the
leading contribution to $\psi - \psi(\infty,r)$ comes from the $n=1$
pole in the Green's function, which produces $q = 2 + g_- = 2$. Thus,
our analytical results are in full agreement with the BCKL numerical
results. 

\subsection{Analytical results for SdS spacetime}

Finding the Green's function for a scalar field in SdS spacetime is
considerably more difficult than what was accomplished in this
section, and this calculation will be the subject of a separate
publication \cite{18}. The conclusion is that for SdS spacetime, the
late-time behavior of the wave function is still exponential, $\psi
\sim e^{-p\kappa_c t}$, with the decay constant $p$ given by the same 
expression as before, apart from corrections of the order of
$r_e/r_c$: 
\begin{equation}
p = l + \frac{3}{2} - \frac{1}{2}\sqrt{9-48\xi} + 
O\biggl( \frac{r_e}{r_c} \biggr).
\label{4.28}
\end{equation}
This is the result that was quoted in Eqs.~(\ref{3.2}) and
(\ref{3.3}). 

In practice, the correction term of order $r_e/r_c$ is too small to be
revealed numerically. Nevertheless,  its presence is important, because it
prevents $p$ from ever becoming an integer. This means that the
cancellation of poles at integer values of $i\omega$, discussed in
Sec.~IV D, is a phenomenon that occurs only for pure de Sitter
spacetime; pole cancellation does {\it not} occur in full SdS
spacetime. Consequently, the cases $\xi = 0$ and $\xi = \frac{1}{6}$
are not exceptional in SdS spacetime, and such qualitative changes of
behavior as depicted in Fig.~5 do not occur.

We have numerically tested the validity of Eq.~(\ref{4.27}) for SdS
spacetime. As expected, the accuracy of this formula is worse than for
pure de Sitter spacetime: With $r_e = 1$, $r_c = 500$, $\sigma = 10$,
and evaluating the wave function at $r^* = 100$, we find that the
error is less than 5\% if $100 < v_c < 200$, but that it climbs to
approximately 15\% when $v_c \sim 50$ or $v_c \sim 400$. 

\section{Electromagnetic and gravitational radiation}

The scalar-wave equation (\ref{1.1}) serves as a model for
the propagation of electromagnetic and gravitational waves in the SdS
spacetime. In this section we indicate how our results can be 
generalized to cover these physically important cases; a more
complete discussion will be found in Ref.~\cite{18}. 

It is well known that electromagnetic and gravitational perturbations 
of the SdS spacetime can be analyzed in terms of a master potential
$\Phi$ which is related to the components of the perturbing 
fields \cite{24}. After a decomposition in spherical harmonics, the
wave function $\psi_l(t,r)$ is found to obey an equation of the
form of Eq.~(\ref{2.13}), but with the generalized potential 
\begin{equation}
V_l(r) = f \biggl[ \frac{l(l+1)}{r^2} -
             \frac{2(s^2-1)M}{r^3} \biggr], 
\label{5.1}
\end{equation}
where $s=1$ ($s=2$) for electromagnetic (gravitational)
perturbations. In both cases the multipole index $l$ is 
restricted by $l \geq s$. 

The considerations of the preceding sections indicate that the late-time
behavior of $\psi_l$ is insensitive to the presence of $M$-dependent terms in
Eq.~(\ref{5.1}).  Removing these terms, we obtain the potential of
Eq.~(\ref{4.2}) with $\xi = \frac{1}{6}$. On these grounds, we expect
that the late-time behavior of $\Phi$ is exponential, $\Phi \sim e^{-p\kappa_c
t}$, with $p$ given by Eq.~(\ref{1.2}) with $\xi = \frac{1}{6}$:
\begin{equation}
p = l + 1 + O\biggl( \frac{r_e}{r_c} \biggr).
\label{5.2}
\end{equation}
This result is borne out in our numerical simulations. 
  
\section*{Acknowledgments}

P.R.B. was supported by the Sherman Fairchild Foundation, and NSF Grants
AST-9417371 and PHY94-07194.  C.M.C. is grateful to the Royal
Commission for the Exhibition of 1851 for financial support. W.G.L and
E.P are supported by the Natural Sciences and Engineering Research
Council of Canada. E.P. wishes to thank Kip Thorne for his kind
hospitality at the California Institute of Technology, where part of
this work was carried out.


\begin{references}
\bibitem{XX}
These integrations are much easier to do, owing to the fact that for de
Sitter spacetime, the relation $r^* = r_c\, \mbox{arctanh}(r/r_c)$ can 
be explicitly inverted. On the other hand, special care must be taken 
with the regular center $r = 0$: The boundary condition 
$\psi_l(t,r=0) = 0$ must be imposed, ensuring the regularity of the 
scalar field $\Phi$ at $r=0$.
\bibitem{foot1} 
More precisely, the overdot on $g$ denotes
differentiation with respect to $t$, while the overdot on $\psi$
denotes differentiation with respect to $t'$, which is set to zero in
Eq.~(4.4). The fact that $g$ depends on $t-t'$ explains why the two
terms within the square brackets are added instead of subtracted.
\bibitem{foot2} 
It should be noted that if $\xi > \xi_c$, the two
poles at $i\omega = l + \frac{3}{2} \pm \frac{i}{2}\sqrt{48\xi - 9}$
are equally near to the real axis, so that both contributions should
be taken into account. We leave this subtlety aside, as it does not
affect the following discussion.
\end{references}
\end{document}